\documentclass[aps,prl,twocolumn,showpacs,amsmath,amssymb,floatfix]{revtex4-1}
\usepackage[utf8]{inputenc}
\usepackage[T1]{fontenc}
\usepackage[english]{babel}
\usepackage[caption=false]{subfig}
\usepackage{lmodern}
\usepackage[usenames,dvipsnames,svgnames,table]{xcolor}
\usepackage{graphicx}
\graphicspath{{img/} }
\usepackage{hyperref}
\usepackage{bm}
\usepackage{color}
\usepackage[utf8]{inputenc} 
\usepackage{cmap}
\definecolor{rltred}{rgb}{0.75,0,0}
\definecolor{rltgreen}{rgb}{0,0.5,0}
\definecolor{rltblue}{rgb}{0,0,0.75}
\hypersetup{colorlinks,%
hypertexnames=true,%
pdfauthor={Stephan Burkhardt, Matthias Liertzer, Dmitry O. Krimer,
  Stefan Rotter},%
pdftitle={Steady-state ab-initio laser theory for lasers with fully or
nearly degenerate resonances},%
pdfview=FitH,%
pdfstartview=FitH,%
pdfpagemode=UseNone,%
bookmarksopen=true,%
bookmarksnumbered=true,%
pdfhighlight=/I,%
linkcolor=rltred,%
citecolor=rltgreen,%
linktocpage=true}

\begin{document}

\title{Steady-state ab-initio laser theory for lasers\\with fully or
  nearly degenerate resonator modes}
\author{Stephan Burkhardt}
\affiliation{Vienna University of Technology, Wiedner Hauptstraße 8--10/136, A--1040 Vienna, Austria, EU}

\author{Matthias Liertzer}
\affiliation{Vienna University of Technology, Wiedner Hauptstraße 8--10/136, A--1040 Vienna, Austria, EU}

\author{Dmitry O.~Krimer}
\affiliation{Vienna University of Technology, Wiedner Hauptstraße 8--10/136, A--1040 Vienna, Austria, EU}
\author{Stefan Rotter}
\affiliation{Vienna University of Technology, Wiedner Hauptstraße 8--10/136, A--1040 Vienna, Austria, EU}

\date{\today}

\begin{abstract}
  We investigate the range of validity of the recently developed
  steady-state ab-initio laser theory (SALT). While very efficient in
  describing various microlasers, SALT is conventionally believed not
  to be applicable to lasers featuring fully or nearly degenerate
  pairs of resonator modes above the lasing threshold.  Here we
  demonstrate how SALT can indeed be extended to describe such cases
  as well, with the effect that we significantly broaden the theory's
  scope. In particular, we show how to use SALT in conjunction with a
  linear stability analysis to obtain stable single-mode lasing
  solutions that involve a degenerate mode pair. Our flexible and
  efficient approach is tested on one-dimensional ring lasers as well
  as on two-dimensional microdisk lasers with broken symmetry.
\end{abstract}


\maketitle
\section{Introduction}
Microcavity lasers are essential elements in modern photonics and have
been realized with cavities of very different shape and with various
lasing
mechanisms~\cite{vahala_optical_2003,cao_dielectric_2015,nockel_ray_1997,gmachl_high-power_1998,harayama_asymmetric_2003,lebental_highly_directional_2006,song_chaotic_2009,wang_whispering-gallery_2010,yang_pump_induced_2010,albert_directional_2012,peng_loss-induced_2014,brandstetter_reversing_2014-1}. The
nonlinear lasing behavior of these systems can, in principle, be
modeled using the semi-classical Maxwell-Bloch (MB)
equations~\cite{haken_nonlinear_1963,lamb_theory_1964,lang_why_1973,haken_laser_1986}. Due
to their time-dependent nature these equations are, however, usually
difficult to solve for all but the most simple cases. In recent years,
a much more efficient approach named steady state ab initio lasing
theory (SALT) has emerged, which can be used to describe the
steady-state lasing of
lasers~\cite{tureci_self-consistent_2006,ge_quantitative_2008,tureci_ab_2009,ge_steady-state_2010,cerjan_injected_2014,esterhazy_scalable_2014,pick_linewidth_2015}.
Among other advances, this new framework has shed light on
weakly-scattering random lasers~\cite{tureci_strong_2008}, on
pump-induced exceptional
points~\cite{liertzer_pump-induced_2012,chitsazi_experimental_2014,brandstetter_reversing_2014-1,peng_loss-induced_2014}
and on coherent perfect
absorption~\cite{chong_coherent_2010,wan_time-reversed_2011} and has
opened up new ways of controlling the emission patterns of random as
well as of microcavity
lasers~\cite{hisch_pump-controlled_2013,liew_control_2014_apl}. One of
the major drawbacks of SALT is that its conventional formulation fails
for the simulation of microlasers with nearly degenerate modes as
occurring, e.g., in whispering gallery mode resonators with an
inherent
symmetry~\cite{lebental_highly_directional_2006,albert_directional_2012,wang_whispering-gallery_2010,gagnon_photonic_molecules_2014,brandstetter_reversing_2014-1,cao_dielectric_2015}.

Here, we present an approach to generalize SALT to such cases. This
extension allows us to observe, among other phenomena, that nearly
degenerate modes may merge into a single-mode. These steady-state
solutions are, however, not necessarily stable with respect to small
time-dependent spatial perturbations. In particular, it has already
been shown, that in highly symmetric systems such as ring lasers not
every steady state solution of the MB equations is necessarily
stable~\cite{zeghlache_bidirectional_1988,risken_instability_1968}. Hence,
the stability of the solutions obtained from our extended SALT
approach has to be verified. Up to now such additional stability
checks were always done using direct time-dependent simulations of the
MB
equations~\cite{ge_quantitative_2008,chua_low-threshold_2011,cerjan_steady-state_2012,liertzer_pump-induced_2012}.
One of the reasons for using SALT, however, is exactly to avoid this
kind of computationally very demanding numerics.

In this work, we thus introduce a much more efficient way to determine
the stability of the SALT solutions based on a rigorous linear
stability analysis for single-mode steady-state solutions.
Furthermore, our work extends the scope of SALT to previously
inaccessible parameter regimes, where even bifurcating solutions
(stable or unstable) of the nonlinear equations can now be
appropriately dealt with.

\section{Short review of SALT}
\label{sec:maxwell-bloch-salt}
In semi-classical laser theory, the dynamics of a laser is governed by
the interaction of classical fields with an ensemble of two-level
atoms as described by the so-called Maxwell-Bloch (MB)
equations. Restricting the fields to 1D or to the transverse magnetic
(TM) polarization in 2D, the electric field and polarization become
scalar~\footnote{The restriction to TM modes is only due to the fact
  that this simplifies the equations. However, the results presented
  in this work equally apply if one uses the full three-dimensional
  MB, where $E^+,P^+$ are vector quantities.}. Using the rotating wave
approximation, the MB equations can be derived as
follows~\cite{haken_laser_1986}
\begin{align}
  \label{eq:mwbequations}
  \begin{split}
    \epsilon \ddot E ^+ &= \nabla^2 E ^+ - \ddot P ^+ \\
    \dot P ^+ &= - \left( i \omega_a + \gamma_{\bot} \right) P^+ - i \gamma_{\bot} E^+ D \\
      \frac{\dot D}{\gamma_{\parallel}} &= D_0 - D + \frac{i}{2}
      \left( E^+ {(P^+)}^* - \mathrm{c.c.} \right)\,.
  \end{split}
\end{align}
In these non-linear partial differential equations, $E^+(x,t)$ and
$P^+(x,t)$ denote the positive frequency components of the electrical
field and the polarization of the medium, respectively. The quantity
$D(x,t)$ is the population inversion of the two-level atoms and
$D_0(x)$ stands for the externally imposed pump strength. The
constants in the MB equations describe the properties of the cavity
and of the gain material: the dielectric function $\epsilon(x)$, the
transition frequency of the two level atoms $\omega_a$, as well as the
decay rates of the polarization $\gamma_{\bot}$ and of the population
inversion $\gamma_{\parallel}$. The boundary conditions for the
equations above are typically outgoing boundary conditions, which
numerically can, e.g., be implemented using a perfectly matched
layer~\cite{berenger_perfectly_1994,esterhazy_scalable_2014}.

The natural units in Eqs.~(\ref{eq:mwbequations}) and all example
systems in this work can be converted back to SI units by choosing an
appropriate length scale $\tilde L$, multiplying all lengths in the
example systems by this quantity $\tilde L$, multiplying the
quantities
($\frac{\partial}{\partial
  t},\gamma_{\parallel},\gamma_{\bot},\omega_a$)
by $c / \tilde L$, where $c$ is the speed of light, dividing
$\frac{\partial}{\partial x}$ by $\tilde L$ and finally by
transforming $E$, $P$, and $D$ as follows
$E ^+ _{\mathrm{SI}} =E^+ \frac{2g}{\hbar \sqrt{\gamma_{\parallel}
    \gamma_{\bot}}}$,
$P ^+ _{\mathrm{SI}} = P^{+} \frac{2 g}{\hbar \epsilon_0
  \sqrt{\gamma_{\parallel} \gamma_{\bot}}}$,
and $D_{\mathrm{SI}} = D \frac{g^2}{\hbar \gamma_{\bot} \epsilon_0}$.
The variable $g$ is the transition dipole moment of the two-level
atoms.

For microlaser systems lasing in steady-state, only a finite number of
modes in the system are active. This is the case described by SALT, in
which the MB equations are simplified to a set of time-independent,
non-Hermitian, nonlinear and coupled Helmholtz equations. The
solutions of these SALT equations are much more efficiently calculated
than those of the MB
equations~\cite{ge_quantitative_2008,chua_low-threshold_2011,cerjan_steady-state_2012,liertzer_pump-induced_2012}.

The cornerstone of the SALT equations is a multi-periodic ansatz for
the electro-magnetic field as well as for the polarization
\begin{equation}
  \label{eq:multiperiodic}
  \begin{split}
    E^{+}(x,t) &= \sum_{\mu=1} ^N E_{\mu}(x) \ e^{-i\omega_{\mu}t}\\
    P^{+}(x,t) &= \sum_{\mu=1} ^N P_{\mu}(x) \ e^{-i\omega_{\mu}t},
  \end{split}
\end{equation}
where each triplet $(E_{\mu}, P_{\mu},\omega_{\mu})$ represents a mode
$E_\mu$ of the system lasing at the \emph{real} frequency
$\omega_{\mu}$. Inserting the ansatz~\eqref{eq:multiperiodic} into the
last equation of~\eqref{eq:mwbequations} results in
\begin{equation}
  \label{eq:beatingincluded}
  \dot D = \gamma_{\parallel} \left( D_0 - D
    \right) \\
     + \frac{i \gamma_{\parallel}}{2} \sum_{\mu,\nu} (E_\mu P^*_\nu e^{i(\omega_\nu-\omega_\mu)t} - \mathrm{c.c.})
\end{equation}
with terms of the form $P_\nu E_\mu ^* e^{i(\omega_\nu-\omega_\mu)t}$,
which explicitly depend on time. For multiple lasing modes, $D$ will
thus never be completely static. However, if the timescale
$1 / (\omega_\nu - \omega_\mu)$ on which these terms oscillate is much
shorter than the timescale on which $D$ varies ($1/\gamma_\parallel$),
their contribution can be neglected. In other words, for systems where
$|\omega_\nu - \omega_\mu| \gg \gamma_{\parallel}$ holds for all pairs
$\mu,\nu$ of active modes~\footnote{As discussed
  in~\cite{esterhazy_scalable_2014} further conditions come into play
  for the ``bad cavity limit'', which, however, we do not consider for
  the systems shown in this paper.} the inversion can be approximated
to be
stationary~\cite{tureci_self-consistent_2006,esterhazy_scalable_2014}.

This stationary inversion approximation (SIA) is, however, not well
satisfied in macroscopic lasers with a large density of modes as well
as in microcavity lasers with an inherent symmetry. We will focus here
on the latter case and shall consider ring- or microdisk lasers with
degenerate modes or slightly perturbed versions of these systems with
nearly degenerate modes. The frequency splitting of these nearly
degenerate modes will typically violate the condition
$\Delta \omega \gg \gamma_{\parallel}$ for realistic values of
$\gamma_{\parallel}$. Since not only the lasing frequencies of these
modes are very close to each other but also their thresholds are at
comparable pump strengths $D_0$, the traditional SALT algorithm will
not be applicable when both modes of such a pair move across the
lasing threshold. For the completely degenerate case this problem is
even more acute.

In the following, we will provide an extension of the SALT algorithm,
which is able to overcome this significant drawback. It does so by
taking into account that degenerate modes with a fixed relative phase
can be expressed as a single active lasing mode within the SALT
formalism.  For nearly degenerate modes, we find that such a mode pair
can become dynamically stable in the form of a single lasing mode,
allowing us to treat the solution again with the above SALT ansatz. In
order to present our approach as clearly as possible, we will focus
here on the case of single-mode lasing only, keeping in mind that our
algorithm can be generalized to multiple pairs of degenerate modes or
multiple modes in general~\cite{burkhardt_stability_2015}.

For a single lasing mode the ansatz~(\ref{eq:multiperiodic}) satisfies
the MB Eqs.~(\ref{eq:mwbequations}) exactly, leading to the following
single-mode SALT equation for the electric field $E_1$ and for the
\emph{real} lasing frequency $\omega_1$:
\begin{equation}
  \Bigg[ \nabla^2 + \left(
    \epsilon(x) +  \Gamma_1\,\frac{D_0(x)}{1+|\Gamma_1 E_1(x)|^2}
  \right) \omega_1^2 \Bigg] E_1(x) = 0\,,
  \label{eq:salt1}
\end{equation}
where
\begin{equation}
  \Gamma_1 = \frac{\gamma_{\bot}}{\omega_1 - \omega_a + i \gamma_{\bot}}\,,
\end{equation}
and the boundary conditions are the same as for the MB
equations. Note, that Eq.~\eqref{eq:salt1} depends nonlinearly both on
the frequency $\omega_1$ as well as on the shape of the mode $E_1(x)$
through a self-saturation spatial hole burning interaction. It can be
straightforwardly solved using a Newton-Raphson solver as described in
detail in~\cite{esterhazy_scalable_2014}. The solver requires an
initial guess which can be obtained by tracking the modes in the
system from an initial value at zero pump strength up to the pump
strength of interest.  This procedure has the advantage that for small
pump values (below the lasing threshold) one can use the following
eigenvalue problem that is linear with respect to the mode profiles,
\begin{equation}
  \Bigg[ \nabla^2 + \Big(
    \epsilon(x) +  \bar{\Gamma}_i\,D_0(x)
  \Big) \bar{\omega}_i^2 \Bigg] \bar{E}_i(x) = 0\,,
  \label{eq:salt_inactive1}
\end{equation}
where
$\bar{\Gamma}_i=\gamma_\bot/(\bar{\omega}_i - \omega_a + i
\gamma_\bot)$.
(Note that we label all quantities below the lasing threshold by
overbars.)  For pump strength $D_0=0$ the complex eigenvalues
$\bar{\omega}_i$ have a negative imaginary part. When increasing the
pump strength, the eigenfrequencies will typically move towards the
real axis (interesting exceptions to this rule are discussed
in~\cite{liertzer_pump-induced_2012,brandstetter_reversing_2014-1,peng_loss-induced_2014}).
Once the first eigenvalue $\bar{\omega}_{1}$ crosses the real axis (we
assume its index $i$ is $1$) it can be used as a guess for solving
Eq.~(\ref{eq:salt1}) for the first lasing mode $E_1$ with real
frequency ${\omega}_1$. This solution can then be tracked to even
higher pump strengths beyond the lasing threshold by repeatedly
solving Eq.~(\ref{eq:salt1}) for increasing $D_0$ while using the
solution from the previous step as an initial guess. Alternative
methods to solve Eq.~(\ref{eq:salt1}) based on an expansion of laser
modes in a bi-orthogonal basis of ``constant-flux states''
\cite{tureci_self-consistent_2006} work in a similar way but will not
be considered here.

In the traditional SALT algorithm the validity of a single-mode
solution $\{E_1, \omega_1\}$ of Eq.~\eqref{eq:salt1} is only
indirectly assessed by keeping track of the remaining passive modes of
the system. While only one mode $\{E_1, \omega_1\}$ is lasing all
these other modes (with $i\neq 1$) have to solve the following
nonlinear, non-Hermitian eigenvalue problem
\begin{equation} \Bigg[ \nabla^2 +
  \left( \epsilon(x) + \bar{\Gamma}_i\,\frac{D_0(x)}{1+|\Gamma_1
      E_1(x)|^2} \right) \bar{\omega}_i^2 \Bigg] \bar{E}_i(x) = 0\,.
  \label{eq:salt_eig}
\end{equation}
The single-mode solution is only valid as long as all other eigenmodes
$\bar{E}_i$ have an eigenvalue $\bar{\omega}_i$ with imaginary
part less or equal to 0, i.e.\@,
\begin{equation}
  \forall i\neq 1: \quad \text{Im}(\bar{\omega}_i) \leq 0\quad\,.
  \label{eq:salt_stab}
\end{equation}
If, however, any other of these eigenvalues crosses the real axis, the
corresponding eigenmode is assumed to be active and incorporated as an
additional lasing mode into the active SALT
equations~\cite{esterhazy_scalable_2014}. As long as the presence of
this second lasing mode does not violate the SIA, the corresponding
two-mode lasing solution is considered stable (as was previously
verified using FDTD
simulations~\cite{ge_quantitative_2008,chua_low-threshold_2011,cerjan_steady-state_2012}).

As we will demonstrate through a comparison to time-dependent
solutions of the MB equations, this simple criterion can not be
applied for nearly degenerate modes. In particular, we show that
stable single-mode solutions may exist even though one of the other
eigenmodes in the system features an eigenvalue $\bar \omega_{i}$ with
positive imaginary part. In order to be able to correctly determine
the stability of a mode when using the SALT Eq.~(\ref{eq:salt1}), we
introduce below a rigorous stability criterion based on a linear
stability analysis.

One of the consequences of this new strategy is that eigenmodes have
to be continuously tracked even when their eigenvalues cross the real
axis without, however, including them as an active lasing
solution. Doing this, we find that these modes exhibit complicated
frequency shifts and bifurcations when varying the pump strength
$D_0$. Examples of this kind will be discussed in the subsequent
sections.

\section{Example 1: symmetric 1D ring laser}
\begin{figure}[ht!]
    \raisebox{1.05cm}{\includegraphics{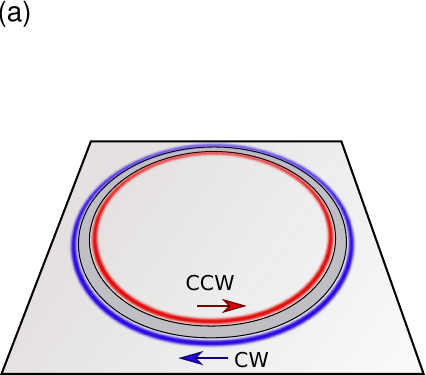}}
    \includegraphics[width=0.48\linewidth]{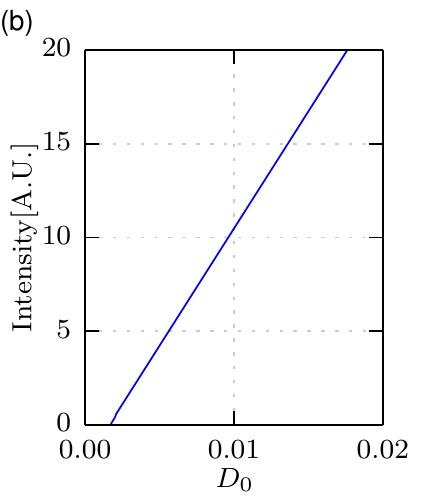}
\\
    \includegraphics{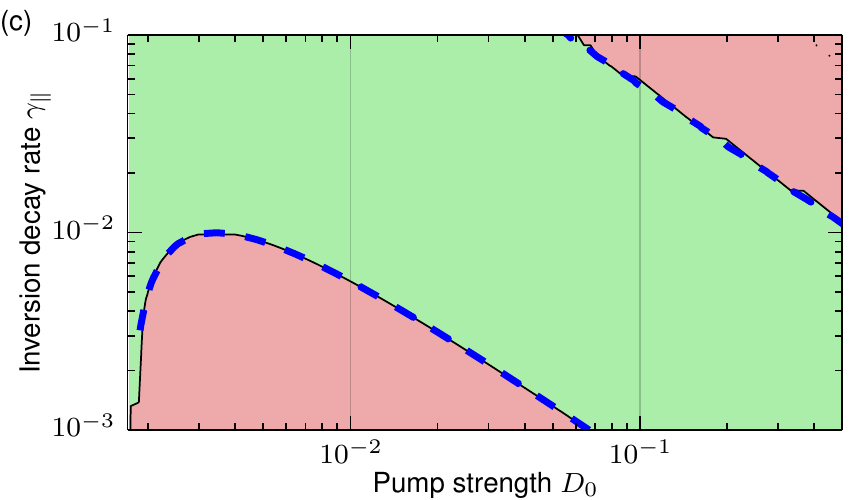}
    \caption{(Color online) Results for a one-dimensional ring laser
      system with circumference $L=1$, $\gamma_{\bot}=1$ and
      $\omega_a=61$. The openness of the system is modeled by a
      lossy dielectric function $\sqrt{\epsilon(x)} =
      1+0.0002i$.
      (a)~Sketch of the ring laser system, which can accommodate
      clockwise (CW) or counterclockwise (CCW) traveling
      modes. (b)~Lasing mode intensity versus applied pump strength
      for the traveling wave solution. Note that the gain parameters have been
      chosen such that only a single pair of degenerate modes reaches
      the lasing threshold. The traveling wave solution in which the
      system lases has a frequency of $\omega \approx 62.65$ which
      stays constant while the pump strength is
      increased. (c)~Stability analysis of the SALT results. Shown are
      the stability of this traveling wave solution under variation of
      the pump strength $D_0$ and the relaxation rate of the inversion
      $\gamma_\parallel$. The results of the FDTD simulation are
      color-coded with green (light gray) marking parameter
      combinations where the SALT solution is stable, whereas the red
      (dark gray) region represents unstable behavior. Blue dashed
      lines show the independent results of the linear stability
      analysis, which provides an excellent estimate for the border
      between stable and unstable regions.}
\label{fig:degenerateRingLaser}
\end{figure}

To convey an understanding of how a lasing system with degenerate
modes can be described in the SALT framework, let us first consider
the well-known example of a rotationally symmetric ring laser whose
solution can become unstable in certain parameter regimes~\cite{risken_instability_1968,lugiato_exact_1986,zeghlache_bidirectional_1988,moloney_fixed_1987,grynberg_observation_1988,giusfredi_optical_1988,lugiato_spontaneous_1989,tureci_mode_competition_2005,vandersande_two-dimensional_2008,sunada_random_2011,takougangkingni_direct-modulation_2012}. We
model the system as a one-dimensional, homogeneous medium with
periodic boundary conditions (see Fig.~\ref{fig:degenerateRingLaser}a
for an illustration). To incorporate losses through absorption and
outcoupling, we set the index of refraction to a complex value.

Solving Eq.~(\ref{eq:salt_inactive1}) for the unpumped system produces
a set of eigenstates $\{\bar \omega_i, \bar E_i\}$, where a
two-dimensional eigenspace is associated with every eigenvalue
$\bar \omega_i$ due to the rotational symmetry of the system. This
eigenspace contains standing waves of the form
$e^{i \bar \omega_i x} \pm \mathrm{c.c.}$ as well as traveling waves
of the form $e^{\pm i \bar \omega_i x}$. While these two pairs of
states as well as their superpositions solve
Eq.~(\ref{eq:salt_inactive1}) below threshold, the non-linear spatial
hole-burning term in the SALT Eq.~(\ref{eq:salt1}) prevents arbitrary
superpositions from being valid solutions above the threshold. This
leaves only two possible steady-state lasing solutions of the
ringlaser at $\bar \omega_i$, corresponding to the well known
clockwise and counterclockwise traveling wave states. However, from
the literature it is known that ring lasers show complex behavior,
including the fact that these traveling wave solutions are not always
stable~\cite{lugiato_exact_1986,risken_instability_1968,zeghlache_bidirectional_1988,moloney_fixed_1987,grynberg_observation_1988,giusfredi_optical_1988,lugiato_spontaneous_1989,tureci_mode_competition_2005,vandersande_two-dimensional_2008,sunada_random_2011,takougangkingni_direct-modulation_2012}.

Our goal here will be to find the single-mode solutions with SALT and
to identify the corresponding regions of stability. To approach this
problem first in the most general way (i.e., independently of the
employed SALT approach), we set up a finite-difference time-domain
(FDTD) method based on a Yee
lattice~\cite{yee_antennas_1966,bidegaray_time_2003}. This tool allows
us to solve the MB equations Eq.~(\ref{eq:mwbequations}) directly,
including the full temporal evolution starting from an initial
distribution of the electric field. Using this approach we first
confirmed that in the single-mode regime indeed only the traveling
modes are stable in certain parameter regimes.  To assess the latter,
the system was initialized in the traveling-wave solution obtained
from the SALT Eq.~(\ref{eq:salt1})~\footnote{We alternatively analyzed
  the situation when the simulation was not initialized in the state
  obtained from the SALT solution but in an arbitrary initial state
  instead. If the parameters of the system were such that the SALT
  solution was stable (see Fig.~\ref{fig:degenerateRingLaser}), the
  system would usually converge to the SALT solution over time. In the
  region close to the border between stable and unstable solution, the
  system only converged to the SALT solution for certain initial
  conditions, hinting at the existence of a second stable
  non-steady-state solution.} and then left to evolve for a certain
amount of time. We analyzed whether at later times the system remained
in the same steady-state solution as at the beginning of the
simulation. If the system stayed in the same state (and did not show
any signatures of deviating from this state), we considered the
solution to be stable.

The stability diagram resulting from these FDTD simulations is shown
in Fig.~\ref{fig:degenerateRingLaser}c. We find that the stability of
the single-mode SALT solutions not only depends on the external pump
strength $D_0$, but also on the inversion decay rate
$\gamma_{\parallel}$.  This finding concurs with previous
results~\cite{zeghlache_bidirectional_1988} and shows that while the
single-mode SALT solution is an exact solution of the MB equations, it
is not necessarily a stable one.  For systems with degenerate passive
modes (as well as with closely spaced modes
$|\Delta \omega| < \gamma_{\parallel}$ discussed below), the stability
of any solution obtained from SALT is therefore not guaranteed and
needs to be independently verified. Using FDTD for such a
verification, as we did above, is, however, much too costly from a
numerical point of view, in particular, as this would nullify the
computational advantages of SALT.  Also the stability analysis
presented in \cite{zeghlache_bidirectional_1988} for rotationally
symmetric ring lasers is rather limited in scope, as it relies on the
availability of exact analytic solutions for the passive modes. We
will thus develop below a more general and rigorous framework to
analyze the stability of SALT solutions that should be generally
applicable.

\section{Linear stability analysis}
\label{sec:linstabana}
This section contains an overview of a linear stability analysis for
solutions of SALT Eq.~(\ref{eq:salt1}) (a full derivation can be found
in appendix A).  Our starting point is to linearize the original MB
equations (\ref{eq:mwbequations}) around the SALT solution and to
assess whether it is stable against small perturbations. In what
follows we concentrate on the single-mode solutions only and thus
insert the following expressions into the MB equations
(\ref{eq:mwbequations}),
\begin{equation}
  \label{eq:salt-solution-form}
  \begin{split}
    E(x,t) &= (E_1(x)+\delta E(x,t)) e^{-i\omega_1 t},\\
    P(x,t) &= (P_1(x)+\delta P(x,t)) e^{-i\omega_1 t},\\
    D(x,t) &= (D(x) + \delta D(x,t)).
  \end{split}
\end{equation}
In this ansatz, $E_1$, $P_1$, $D$ denote, respectively, the electric
field, the polarization and the inversion of the single-mode SALT
solution of Eq.~\eqref{eq:salt1} and $\delta E$, $\delta P$,
$\delta D$ are the corresponding small perturbations around
it. Utilizing the fact that the SALT solution also exactly solves the
MB equations and neglecting the higher-order contributions of the
perturbations, we derive a set of linear PDEs (\ref{eq:pertSALT1})
with respect to $\delta E$, $\delta P$, $\delta D$.

As a next step we convert the resulting system of equations into a
standard eigenvalue problem. A split of the complex variables into
their imaginary and real parts gives rise to a set of linear equations
for five independent fields, which for convenience can be represented
as a single vector field,
$\vec F(x) = (\mathrm{Re}(\delta E), \mathrm{Im}(\delta E),
\mathrm{Re}(\delta P), \mathrm{Im}(\delta P), \delta D)$.
We look for solutions of the form
$\vec F(x,t) = \vec F(x) e^{\sigma t}$, where $\sigma$ is the growth
rate, and derive a set of linear equations (\ref{Eq_lin_stab_spatial})
containing the spatial dependence only. Using an appropriate
discretization scheme and taking into account the periodic boundary
conditions, we finally end up with a quadratic eigenvalue problem of
the following form:
\begin{equation}
A \vec F + \sigma B \vec F +  \sigma^2 C \vec F = 0,
\label{eq:quadraticEV}
\end{equation}
where $A$, $B$ and $C$ are the corresponding matrices whose dimensions
depend on the chosen spatial discretization.

\begin{figure}[ht!]
  \centering
  \includegraphics[width=\linewidth]{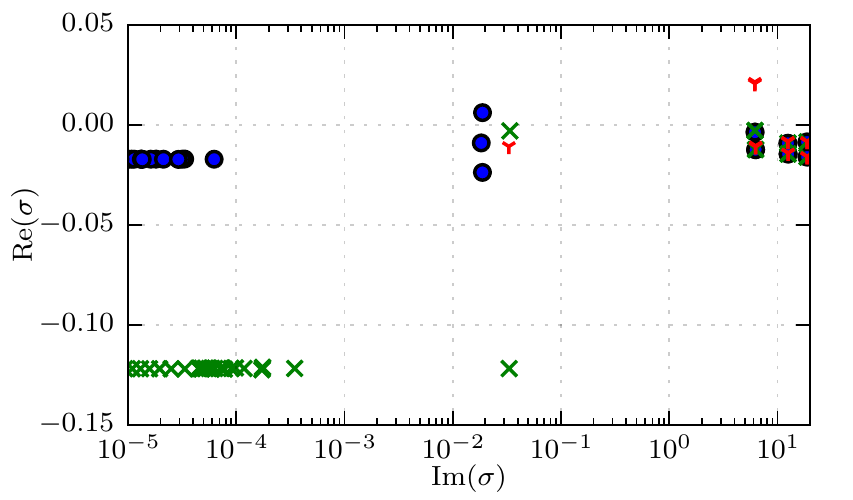}
  \caption{(Color online) Stability eigenvalues $\sigma^j$ of the ring
    laser for $D_0=0.06$ and $\gamma_{\parallel}=10^{-3}$ (blue
    circles), $\gamma_{\parallel}\approx 7\cdot 10^{-3}$ (green
    crosses) and $\gamma_{\parallel}=10^{-1}$ (red Ys). All other
    parameters are chosen as in
    Fig.~\ref{fig:degenerateRingLaser}. One can see that the
    single-mode lasing solution is only stable for the intermediary
    value of $\gamma_{\parallel}$ (green crosses), since for both
    other cases, a $\sigma^j$ with a real part larger than zero
    exists. The stability eigenvalues with
    $\mathrm{Im}(\sigma^i) \approx 10^{-2}$ are related to the
    competition between the nearly degenerate modes. The other visible
    clusters of stability eigenvalues at
    $\mathrm{Re}(\sigma^i)\approx -\gamma_{\parallel}$ are related to
    damped relaxation oscillations observed when initially turning on
    the laser. The eigenvalues with $\mathrm{Im}(\sigma^j) > 10$ are
    related to other resonances of the system that could potentially
    also turn into active lasing modes.}
\label{fig:sigma_plot}
\end{figure}

This eigenvalue problem can be solved numerically, resulting in a set
of eigenvalues and eigenvectors, $\{\sigma^j, F^j(x) \}$. Note that
eigenvalues with $\text{Re}(\sigma^j) > 0$ stand for the
perturbations, which grow exponentially in time implying that our SALT
solution is unstable.  Conversely, if all eigenvalues $\sigma^j$ have
a real part smaller than zero, the SALT solution is stable against
small perturbations. Therefore, finding the eigenvalue with the
largest real part is sufficient to classify the stability of the SALT
solution. The imaginary part of $\sigma$ stands for the frequency
relative to $\omega_1$, with which the perturbation oscillates.  In
Fig.~\ref{fig:sigma_plot}, we show a typical example of the eigenvalue
spectra $\{\sigma^j\}$ with different values of $\gamma_{\parallel}$
for the case of the symmetric ring laser described in the previous
section. Note that due to the fact that the MB equations for the
single-mode regime are invariant under a global phase rotation, the
value $\sigma=0$ always shows up. It, however, does not affect the
behavior of the system and therefore is always excluded from the
consideration.

To assess the stability diagram of a SALT solution we start at a
certain value of the pump strength and then gradually vary the value
of $\gamma_{\parallel}$. The dashed lines in the stability diagram
depicted in Fig.~\ref{fig:degenerateRingLaser}c correspond to the
stability thresholds at which the eigenvalue with the largest real
part, $\max_j[\mathrm{Re}(\sigma^j)]$, crosses the imaginary axis
[$\mathrm{Re}(\sigma)=0$]. We emphasize that the boundaries between
stable and unstable regions, which we find in this way, are in
excellent agreement with the time-dependent simulations, as seen in
Fig.~\ref{fig:degenerateRingLaser}c. It should be noted that previous
studies on such a linear stability analysis involved more restrictive
approximations and a limited class of perturbations by keeping track
of low order Fourier terms only~\cite{zeghlache_bidirectional_1988},
whereas our approach is exact in the framework of the MB equations.

Note, that for the above procedure to work it is not necessary to
compute the whole eigenvalue spectrum $\{\sigma^j\}$, but it is
sufficient to only consider eigenvalues in the complex region close to
the real axis and for imaginary parts in the range of $0$ to
$\omega_1$. In systems where the rotating wave approximation is
justified, only perturbations within this frequency range can
realistically influence the system. In systems with a complex
refractive index, spurious solutions that need to be excluded from the
analysis, can occur in the region $|\mathrm{Im}(\sigma)| > \omega_1$.
This restriction on the eigenvalues we are looking for can be used
together with an iterative eigenvalue solver to check relatively
quickly if a solution is stable or not~\cite{saad_numerical_1992}.

It is worth noting that for the above stability analysis of the MB
equations no additional assumptions have been made and the results are
therefore valid for arbitrary single-mode SALT solutions. In the next
section we consider the ring laser with an embedded scatterer where
modes are only near-degenerate to see how our results change for the
case of such a slight lift of the degeneracy.

\section{Example 2: 1D ring laser with broken symmetry}
\label{sec:exampl-1d-ringl}
\begin{figure}[ht!]
  \centering
    \includegraphics[width=\linewidth]{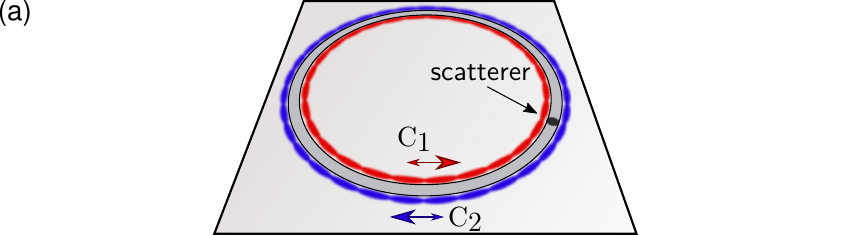} \vspace{0.5em}\\
  \includegraphics{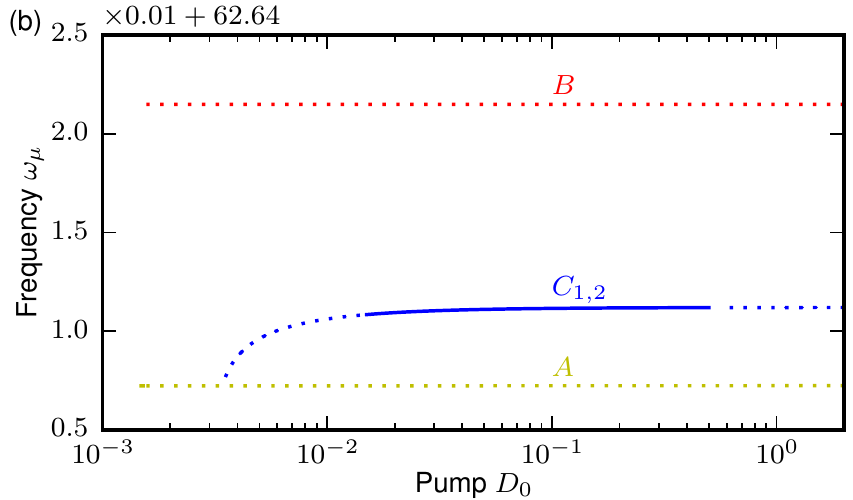}
\\
    \includegraphics[width=\linewidth]{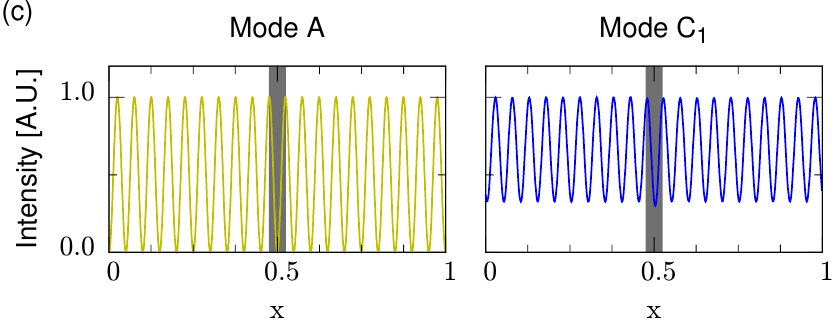}
    \caption{(Color online) Results for a ring laser system as in
      Fig.~\ref{fig:degenerateRingLaser} with
      $\gamma_{\parallel}=0.01$, but with an additional scatterer
      added with $l=0.05$ and
      $\sqrt{\epsilon}=1.05+0.0002i$. (a)~Sketch of the
      system. (b)~Frequencies of single-mode SALT solutions.  It can
      be seen that the two degenerate modes of the symmetric system
      from Fig.~\ref{fig:degenerateRingLaser} have split into modes
      $A$ and $B$. Additionally, at $D_0\approx0.003$, two
      predominantly traveling wave solutions $C_1$ and $C_2$ branch
      off from solution $A$. These two solutions are mirror-symmetric
      images of each other. Dotted lines mark unstable, solid lines
      mark stable solutions. (c)~Intensity distribution of modes $A$
      and $C_1$ with the scatterer marked in gray. Mode $A$ is shown
      directly at threshold ($D_0\approx0.0015$), mode $C_1$ at
      $D_0=0.1$. It can be seen that while $A$ is a standing wave with
      minimal intensity $0$ at the nodes, the predominantly traveling
      wave $C_1$ features a non-vanishing intensity everywhere in the
      cavity. The stability of mode $C_{1,2}$ is studied in
      Fig.~\ref{fig:stability_dedegen}.}
\label{fig:dedegenerateRinglaser}
\end{figure}

In any real-world implementation of a ring laser or microdisk laser,
the rotational symmetry will be slightly broken due to inhomogeneities
in the material or by imperfections of the manufacturing
process~\cite{liew_control_2014_apl,liew_pump-controlled_2014}. While
intuitively one would expect that a slight modification of the system
should not change the structure of the lasing solutions, there is
evidence that this symmetry breaking can have a strong effect on the
lasing
modes~\cite{liew_pump-controlled_2014,wiersig_structure_2011}. In
order to better understand the impact of such a slight symmetry
breaking on the stability of the SALT solutions, we analyze here a
ring laser where the rotational symmetry is broken by a scatterer as
depicted in Fig.~\ref{fig:dedegenerateRinglaser}a.

Since the inhomogeneity in this system breaks the rotational symmetry,
solving Eq.~(\ref{eq:salt_inactive1}) for the inactive system will not
produce any traveling wave solutions. Instead, standing wave solutions
similar to those found in the completely symmetric system are
found. But while the standing wave solutions of the symmetric system
always occurred in degenerate pairs, breaking the symmetry lifts the
degeneracy such that the new modes slightly differ in their complex
frequencies $\bar \omega_{i}$. Therefore, they will neither possess
exactly the same lasing frequency, nor will they reach the lasing
threshold at exactly the same same pump strength $D_0$. Since the
frequency splitting does not fulfill the stationary inversion
approximation for realistic values of $\gamma_\parallel$, a possible
two-mode solution can not be described by SALT (at least not without
an explicit stability analysis).
\begin{figure}[ht!]
  \centering
  \includegraphics[width=\linewidth]{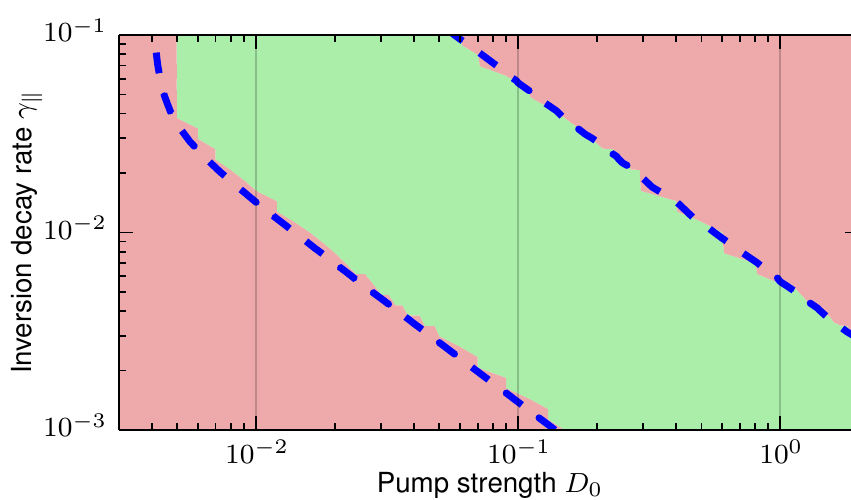}
  \caption{(Color online) Stability of the SALT solution for the
    broken symmetry ring laser shown in
    Fig.~\ref{fig:dedegenerateRinglaser}. The analyzed solution is the
    one marked as branch $C_{1,2}$ in
    Fig.~\ref{fig:dedegenerateRinglaser}b. While the shape of the
    stability region is very similar to the one observed in
    Fig.~\ref{fig:degenerateRingLaser}, the broken symmetry system
    requires a minimal pump strength $D_0$ to compensate for the
    frequency splitting of the nearly degenerate modes before a stable
    predominantly traveling wave solution can emerge.}
\label{fig:stability_dedegen}
\end{figure}

In our analysis we consider gain parameters that only support the
lasing action of a single pair of such modes (here located at
$\bar \omega_i \approx 62$) such that only the single-mode solutions
of SALT seem viable candidates for steady-state lasing. These
single-mode SALT solutions are shown in
Fig.~\ref{fig:dedegenerateRinglaser}b in form of their frequency
dependence on the pump strength $D_0$.  In particular, the two
single-mode solutions of SALT corresponding to the pair of nearly
degenerate resonator modes are shown as mode $A$ and $B$. When
tracking mode $A$ from its threshold while increasing the pump
strength, it shows the following behavior: At $D_0\approx 0.0015$ the
standing wave single-mode solution $A$ becomes active and the system
starts lasing (see Fig.~\ref{fig:dedegenerateRinglaser}b). At this
pump value the mode is stable, both according to the traditional SALT
criterion for stability as well as in the linear stability
calculations. However, a very small increase of the pump strength
brings the eigenvalue $\bar \omega_i$ of the second mode of this
near-degenerate pair to the real axis and thus renders the single-mode
solution unstable (again, according to both criteria).  Whether a
stable two-mode solution in this near-degenerate regime exists is a
question that presently falls outside the scope of SALT (due to the
non-stationary inversion)\cite{burkhardt_stability_2015}.  However,
with the techniques presented above we can investigate the stability
of a single-mode solution in the regime beyond the point where the
second mode passes the instability threshold.  For this purpose, we
track each of the two modes $A$ and $B$ towards higher pump strength
using the SALT equation Eq.~(\ref{eq:salt1}) for each mode as if the
other mode was not active.  Using, in addition, the stability
criterion from the previous section reveals that both solutions on
their own remain unstable for higher values of the pump strength. We
find, however, that a bifurcation occurs at $D_0 \approx 0.003$, at
which two further single-mode SALT solutions $C_{1,2}$ of
Eq.~(\ref{eq:salt1}) branch off from solution $A$. These new modes
share the same frequency, which lies approximately in between the
frequencies of the unstable modes $A$ and $B$.  To find these two
modes one can, e.g.\@, use a linear superposition of mode $A$ and $B$
at higher pump strengths as an initial guess for solving
Eq.~(\ref{eq:salt1}), similarly to the way the traveling wave
solutions of the symmetric ring laser can be expressed as linear
superpositions of the standing wave solutions.

By looking at the mode profiles of solutions $C_{1,2}$ ($C_{1}$ is
shown in Fig.~\ref{fig:dedegenerateRinglaser}c), we observe that these
modes are related to the traveling wave solutions observed in the
system with unbroken rotational symmetry. There, the two stable
solutions were purely clockwise or counter-clockwise traveling
waves. Here, each of the modes $C_{1,2}$ still features a dominant
contribution in one direction and is identical to the other mode when
being reflected at the symmetry axis containing the scatterer
($x=0.5$). The major difference to the solutions observed in the
unbroken ring laser is that the solutions $C_{1,2}$ do not exist below
a critical pump strength $D_{\mathrm{crit}}\approx 0.003$, since the
nonlinear term in Eq.~\eqref{eq:salt1} needs to be strong enough to
compensate for the frequency splitting between the two modes.  The
fact that the nonlinear term is responsible for compensating the
frequency difference between the two nearly degenerate modes was
further investigated. We used a linear approximation to estimate the
frequency shift experienced by the passive mode of the system that
corresponds to the single-mode solution $B$ as a consequence of the
spatial hole burning from the active mode $A$. For this purpose, we
modeled the change in the population inversion $\Delta D(x)$ as well
as the resulting change in frequency $\Delta \bar \omega_{B}$ of the
passive mode as a linear function of the pump strength $D_0$. The
estimate predicted the two modes to share their frequency at a value
of $D_0 \approx 0.30$, which is very close to the point
$D_0\approx0.33$, where the solutions $C_{1,2}$ really emerge. The
remaining small discrepancy can be attributed to the inaccuracy of the
employed approximations.

Next, we check the stability of the mode pair $C_{1,2}$. First of all
we recall that using the traditional SALT criterion
Eq.~\eqref{eq:salt_stab} (which is not applicable here), one would
find that the solutions $C_{1,2}$ are never stable.  However, both the
results of the FDTD calculations as well as the linear stability
calculations do show that these modes are stable in well-defined
limits.  These limits are indicated in
Fig.~\ref{fig:stability_dedegen}, where the stability of the
single-mode solutions are depicted as obtained from both methods under
variation of the pump strength $D_0$ as well as of the relaxation rate
of the inversion $\gamma_\parallel$. As discussed in the above
paragraph, there is a small region of stability for very low values of
pump strength $D_0$ where mode $A$ is stable. Of more interest,
however, is the large region of stable lasing for modes $C_{1,2}$
which opens up at the critical pump strength
$D_{\mathrm{crit}}\approx 0.003$ (green region in
Fig.~\ref{fig:stability_dedegen}) and which is similar to the one
observed for the fully symmetric ring laser
(Fig.~\ref{fig:degenerateRingLaser}).

Altogether, the results of the linear stability analysis are in
excellent agreement with the findings of the time-dependent
simulations of the MB equations, as can be seen both in
Fig.~\ref{fig:degenerateRingLaser}b and in
Fig.~\ref{fig:stability_dedegen}. The linear stability analysis thus
proves to be a reliable tool to classify the stability of SALT
solutions and can therefore be used even for systems where verifying
the results through time-dependent simulations is not feasible. This
is, in particular, the case for higher dimensional systems, such as
the microdisk system analyzed in the next section, where
time-dependent simulations for a large range of parameters is
computationally very demanding.

\section{Example 3: 2D microdisk laser with wedge}
\begin{figure}[ht!]
  \centering
  \includegraphics{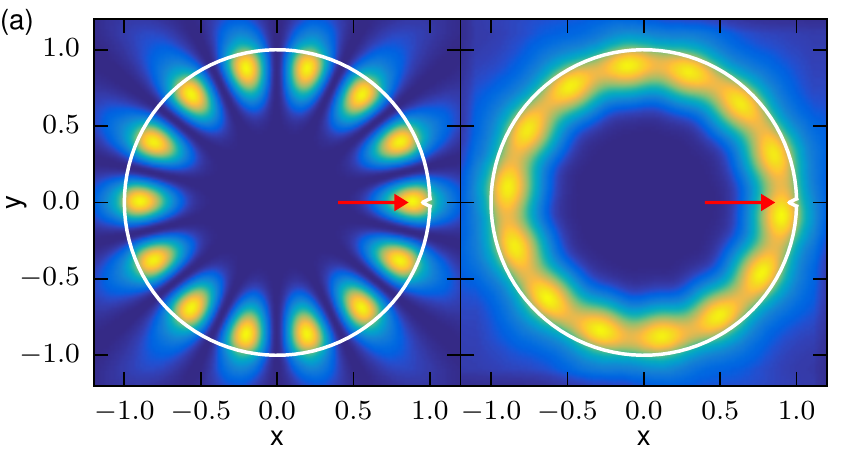}
  \includegraphics{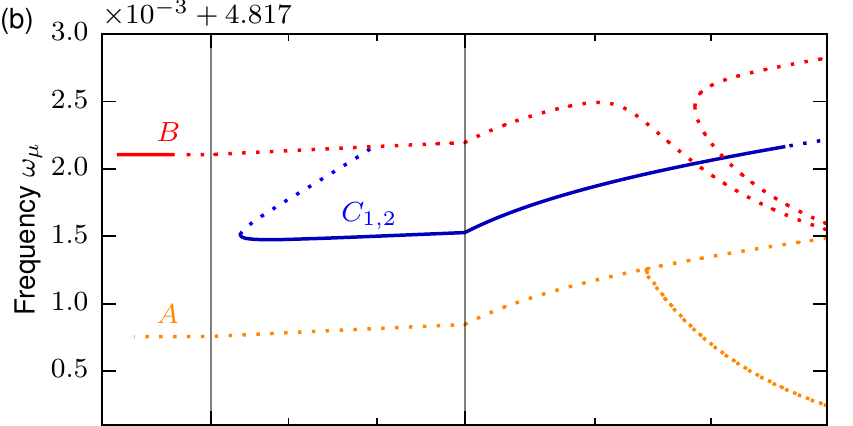}
  \includegraphics{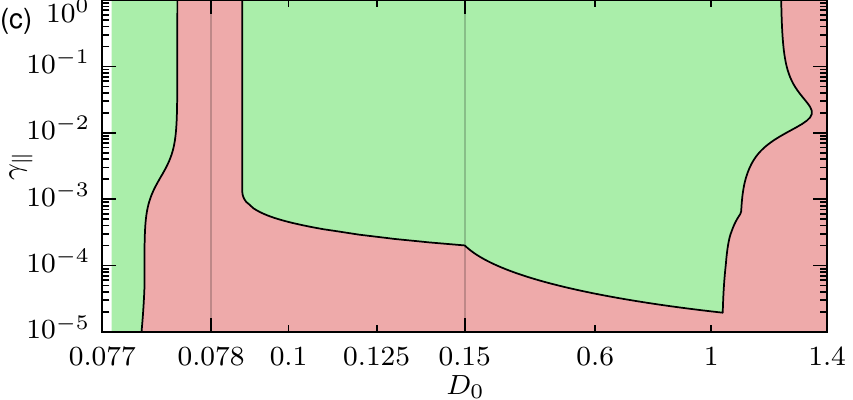}
  \caption{(Color online) Single-mode lasing states for a 2D cavity
    with a wedge. The disk has a radius of $1$, a refractive index of
    $n=2+0.01i$ and the wedge on the right side of the disk has a
    depth of $0.05$ and a width of $0.05$. The gain parameters are
    $\omega_a=4.83$ and $\gamma_\bot=0.1$. (a)~The cavity is outlined
    in white with arrows marking the position of the wedge. The
    spatial intensity pattern of two stable single-mode solutions are
    shown: (left) Standing wave mode $B$ at pump strength $D_0=0.0773$
    with regularly spaced nodes with intensity zero; (right)
    dominantly traveling wave solution $C_1$ at pump strength
    $D_0=0.15$. (b)~Laser frequencies of single-mode solutions to the
    SALT Eq.~(\ref{eq:salt1}). Curves $A$ (orange) and $B$ (red)
    correspond to two standing wave solutions, which have even and odd
    symmetry with respect to the $x$-axis. The two modes represented by
    curve $C_{1,2}$ (blue) feature a broken symmetry, but are
    mirror-symmetric to each other. Solid (dotted) lines denote a
    stable (unstable) solution for $\gamma_\parallel=0.01$. Note, that
    the pump axis is separated into three regions of different linear
    scaling for the sake of clarity. (c)~Stability diagram of the
    single-mode solutions shown in (b) with respect to the pump
    strength $D_0$ and the relaxation rate of the inversion,
    $\gamma_\parallel$. The color-coding is as in
    Fig.~\ref{fig:degenerateRingLaser}c except that the stability is
    here solely determined by the linear stability analysis.}
\label{fig:results_2dwedge}
\end{figure}

Next, we will extend our results to a two-dimensional system, where
full time-dependent simulations become impractical for the required
range of parameters, but the linear stability analysis can still be
easily performed to justify the stability of the SALT solutions. The
system we will consider is a perturbed 2D microdisk laser where the
cavity is slightly deformed by cutting a small wedge into the disk
(see arrow and white outline in Fig.~\ref{fig:results_2dwedge}a). In
analogy to the scatterer for the 1D ring laser, the wedge has the
effect that the two originally degenerate threshold modes of the
microdisk split (here the frequency splitting
$\Delta \omega \approx 1.4\cdot10^{-3}$). Hence, the multi-mode
condition $\gamma_\parallel \ll \Delta \omega$ cannot be satisfied for
reasonable values of $\gamma_\parallel$ such that a two mode solution
does not fulfill the SIA and the traditional SALT algorithm cannot be
applied.

The strategy to find the single-mode SALT solutions is a bit more
sophisticated as compared to the 1D system. We start with the
threshold modes $A$ and $B$ corresponding to the nearly degenerate
resonances of the unpumped system and track them while gradually
increasing the pump strength $D_0$ (see
Fig.~\ref{fig:results_2dwedge}b).  Mode $B$ has a lower threshold and
is stable in a tiny pump region starting from its threshold at
$D_0\approx0.0771$ up to about $0.0777$.  The corresponding stable
parameter region for mode $B$, shown in the stability diagram in
Fig.~\ref{fig:results_2dwedge}c by the green region on the very left
of the figure, depends on both the pump strength $D_0$ and on
$\gamma_\parallel$. After mode $B$ has become unstable the mode is
tracked further (neglecting the presence of mode $A$ whose resonance
eigenvalue has meanwhile also crossed the real axis). At
$D_0\approx 0.124$ a pair of solutions $C_{1,2}$ branches off from
mode $B$. Whereas mode $B$ and mode $A$ feature a perfect even and odd
symmetry with respect to the $x$-axis (i.e., the symmetry axis of the
system), the solutions $C_{1,2}$ do not possess this symmetry, but
rather are mirror images of each other (compare mode profiles of modes
$B$ and $C_1$ in Fig. 5a).  In fact, the symmetry of the system is
spontaneously broken when either of these modes are lasing, a
phenomenon that has been previously observed in simulations as well as
in experiments~\cite{harayama_asymmetric_2003}.

In contrast to mode $B$, mode $A$ is never a stable laser mode since
it has a higher lasing threshold. At $D_0\approx 0.8$ both (unstable)
modes $A$ and $B$ feature two further branches. These can be
understood as follows: Since the wedge of the two dimensional cavity
only represents a small perturbation to the system, the symmetry with
respect to the $y$-axis is only slightly broken, and, hence, modes $A$
and $B$ are nearly symmetric with respect to this axis (see left panel
of Fig.~\ref{fig:results_2dwedge}a for the intensity pattern of mode
$B$). At $D_0\approx0.8$, this near-symmetry is no longer
realized. However, since the modes have never been fully symmetric,
there is not a single point at which the symmetry breaks, but rather a
smooth transition (as compared, e.g., to the symmetry breaking
transition with respect to the $x$-axis at $D_0\approx0.124$). For
mode $B$, this smooth transition is clearly visible in
Fig.~\ref{fig:results_2dwedge}b. For mode $A$ this transition occurs
in a much smaller pump interval, since mode $A$ has a node directly
located at the wedge of the 2D cavity and its symmetry is therefore
only very slightly distorted.

To obtain modes $C_{1,2}$, we need to track them backwards in pump
strength starting at the branching point at $D_0\approx0.124$ (see
Fig.~\ref{fig:results_2dwedge}b). Reducing the pump strength further,
we observe that a sharp bend occurs in the frequency dependence of
these two modes at $D_0\approx0.86$, where each of the modes $C_{1,2}$
has evolved into a dominantly traveling wave mode (see right panel of
Fig.~\ref{fig:results_2dwedge}a), similar to the nearly degenerate 1D
ring laser. Note that the small contribution traveling in clockwise
direction can be observed as a modulation in the intensity pattern.
Beyond this turning point, modes $C_{1,2}$ become stable in a large
region of parameters $D_0$ and $\gamma_\parallel$ as depicted by the
central green area in the stability diagram in
Fig.~\ref{fig:results_2dwedge}c. The easiest way to find the branch
$C_{1,2}$ numerically is to sum the fields of modes $A$ and $B$ with
an additional relative phase of $\pi/2$ at a pump strength
$D_0 > 0.86$ and use this as a guess for the nonlinear solver to
converge towards one of the dominantly traveling wave modes. The
solution can then be tracked to uncover the whole branch of modes
$C_{1,2}$. Using similar superpositions of already known modes as
starting point for the nonlinear solver, we found several more
branches of non-linearly induced single-mode SALT solutions within the
plotted frequency range, albeit none of them are stable (these
branches are not shown in Fig.~\ref{fig:results_2dwedge}c).

The traveling wave solutions $C_{1,2}$ only remain stable until a pump
strength of $D_0\approx1.3$. Here, the single-mode SALT solution
becomes unstable due to the fact that an additional mode would start
to lase. We also find that for values of
$\gamma_\parallel < 2\cdot 10^{-5}$ the modes $C_{1,2}$ are
never stable, which highlights again how important it is to take into
account the value of $\gamma_\parallel$ for assessing the stability of
a SALT mode.  Traditionally in SALT single-mode lasing solutions were
implicitly considered as stable (without considering
$\gamma_\parallel$)~\cite{tureci_self-consistent_2006,esterhazy_scalable_2014},
but these single-mode solutions were always only identified for the
case where only a single one of the eigenvalues $\bar \omega_i$ of
Eq.~(\ref{eq:salt_eig}) has a non-negative imaginary part. Our results
show that single-mode SALT solutions can also exist for the case of
multiple eigenvalues $\bar \omega_i$ featuring a non-negative
imaginary part. In this new situation,
a stability analysis is, however, indispensable for correctly
interpreting the solutions of the SALT equation. When the results of
the stability analysis are taken into account, the SALT equation
allows us to accurately describe the steady-state of these systems
without requiring any time-dependent simulations.

\section{Conclusion}
In this work we demonstrate that SALT can be used to describe the single-mode
lasing regime of resonators with degenerate or near-degenerate mode pairs.
Our approach builds on a careful tracking of SALT modes in the
non-linear lasing regime together with a linear stability analysis to
judge the validity of the resulting solutions.  The accuracy of the stability
analysis itself was tested by a comparison with full time-dependent
simulations based on the Maxwell-Bloch equations, which shows
excellent agreement in all cases.

Our approach is ideally suited to treat microdisk whispering-gallery-mode
lasers, which were previously difficult to simulate with SALT and
often only accessible through time-dependent simulations or through
strongly simplified models. Generally speaking, our work paves the way to study
interesting non-linear phenomena, such as bifurcating solutions
etc.\@ within the efficient framework of SALT\@.

One obvious direction for further study is the generalization of our
stability analysis to systems in the multi-mode lasing regime, which
will be the aim of a subsequent paper~\cite{burkhardt_stability_2015}.

\section{Acknowledgments}
\begin{acknowledgments}
  The authors acknowledge fruitful discussions with Claas Abert, Sofi
  Esterhazy, Thomas Hisch, and Hakan E.\@ Türeci. Financial support
  by the Vienna Science and Technology Fund (WWTF) through Project
  No. MA09-030 (LICOTOLI) and by the Austrian Science Fund (FWF)
  through Project No.~SFB~NextLite F49-P10 is gratefully
  acknowledged. The computational results presented have been achieved 
  using the Vienna Scientific Cluster (VSC).
\end{acknowledgments}

\appendix*
\section{Appendix A:\@ Derivation of the linear stability analysis}
\label{sec:deriv-line-stab}
We start from a solution $\{E_1(x),\omega_1\}$ of the SALT
Eq.~(\ref{eq:salt1}) at a given pump strength $D_0$. From this, one
can construct the polarization, $P_1$, and inversion, $D$, that show
up in the MB Eqs.~(\ref{eq:salt1}) via
\begin{eqnarray}
D(x)&=& \frac{D_0}{1+|\Gamma_1 E_1(x)|^2}\\
P_1(x)&=& \Gamma_1 D(x) E_1(x).
\end{eqnarray}
Using these quantities, we insert the
ansatz~(\ref{eq:salt-solution-form}) into the MB
Eqs.~\ref{eq:mwbequations}. Using the fact that
$(E_1 e^{-i \omega_1 t}, P_1 e^{-i \omega_1 t}, D)$ is a solution of
the MB equations, we linearize the equations with respect to the
perturbations. This results in the following partial differential
equations for the perturbations of the electric field, the
polarization, and the inversion
\begin{widetext}
\begin{eqnarray}
  \label{eq:pertSALT1}
  \epsilon\,\delta \ddot E =& \nabla ^2 \delta E  - \delta \ddot P + \omega_1^2
                             (\delta P + \epsilon\,\delta E)\, +
                              2 i \omega_1 (\delta
                             \dot P + \epsilon\,\delta \dot E)\\
  \label{eq:pertSALT2}
  \delta \dot P =& \left( i (\omega_1 - \omega_a) - \gamma_{\bot} \right) \delta P -
                  i \gamma_{\bot} (E\,\delta D + \delta E\,D) \\
  \label{eq:pertSALT3}
  \delta \dot D =& -\gamma_{\parallel}\,\delta D
                   + \frac{i\gamma_{\parallel}}{2} \big( \delta E \,P_1^* + E_1\,\delta P^* - \delta E^* P_1 - E_1^*\,\delta P \big).
\end{eqnarray}
\end{widetext}
While this system of equations is linear in the perturbation, it
includes the terms $\delta P^*$ and $\delta E^*$, which can not be
expressed as linear combination of $\delta P, \delta E, \delta D$. In
order to produce a completely linear system of equations, we therefore
split the two complex fields and corresponding perturbations, as well
as a possibly complex dielectric function $\varepsilon$ into their
respective real and imaginary parts and consequently split
Eqs.~\eqref{eq:pertSALT1} and \eqref{eq:pertSALT2} into four real
equations. This yields a linear system of five equations with purely
real terms. These correspond to the five independent fields
$\mathrm{Re}(\delta E)$, $\mathrm{Im}(\delta E)$,
$\mathrm{Re}(\delta P)$, $\mathrm{Im}(\delta P)$, $\delta D$, which
for convenience we can summarize as a single vector field $\vec F$.

In order to analyze if a solution $E_1$ of the SALT equation
\eqref{eq:salt1} is stable we need to check if a perturbation exists
which doesn't relax back to the stable solution. Hence, we make an
ansatz of the form $\vec F(x,t) = \vec F(x) e^{\sigma t}$ which yields
\begin{widetext}
  \begin{eqnarray}
  \label{Eq_lin_stab_spatial}
    \nonumber
    (\delta P_r - \epsilon_i\delta E_i
    + \epsilon_r \delta E_r) \sigma^2 +
    2(\delta P_i +  \epsilon_i \delta E_r + \epsilon_r \delta E_i
    )  \omega_1\,\sigma - (\delta P_r - \epsilon_i \delta E_i +  \epsilon_r \delta E_r)\,
    \omega_1^2  -\nabla^2 \delta E_r &= 0\\\nonumber
    (\delta P_i + \epsilon_i \delta E_r + \epsilon_r \delta E_i)
    \sigma^2 - 2 (\delta P_r - \epsilon_i \delta E_i + \epsilon_r
    \delta E_r) \omega_1 \sigma - (\delta P_i + \epsilon_i \delta E_r
    + \epsilon_r \delta E_i) \omega_1^2 - \nabla^2 \delta E_i &= 0\\\label{eq:pert_qev}
    \delta P_r \sigma + \gamma_\bot(\delta P_r - D \delta E_i - \delta
    D E_i^1) + (\omega_1 - \omega_a) \delta P_i &= 0\\\nonumber
    \delta P_i \sigma + \gamma_\bot(\delta P_i  + D \delta E_r +
    \delta D E_r^1) - (\omega_1 - \omega_a) \delta P_r &= 0\\\nonumber
    \delta D \sigma + \gamma_\parallel (\delta D + P_r^1 \delta E_i +
    \delta P_r E_i^1 - \delta E_r P_i^1 - E_r^1 \delta P_i) &\,= 0,    
  \end{eqnarray}
\end{widetext}
where we have abbreviated $\text{Re}(\cdot)$ and $\text{Im}(\cdot)$
through the sub-indices $r$ and $i$, respectively. In addition we need
to impose boundary conditions for the perturbation $\delta E$. For the
periodic 1D ring laser we can simply assume periodic boundary
conditions, i.e.,
$\delta E(x_{\text{left}}) = \delta E(x_{\text{right}})$ and for the
2D system
$\vec \nabla \delta E \overset{r\to\infty}{=} i (\omega_1 - i \sigma)
\delta E\,\hat{e}_r$.
In a next step Eqs.~(\ref{eq:pert_qev}) are discretized using a
suitable discretization scheme \cite{esterhazy_scalable_2014} which
leads to a quadratic eigenvalue problem that can easily be linearized.
In our calculations we have chosen the finite element framework
FEniCS~\cite{logg_automated_2012} for discretizing
Eqs.~\eqref{eq:pert_qev} and used a perfectly matched layer for
imposing the outgoing boundary conditions in the 2D case. For solving
the linearized quadratic eigenvalue problem we have used SLEPc
\cite{hernandez_slepc:_2005}.

\bibliography{draft,additional}
\end{document}